\documentclass[journal]{IEEEtran}
\usepackage{amsmath,graphicx}
\usepackage{comment}
\usepackage{epstopdf}
\usepackage{mathtools}
\usepackage{mathrsfs}
\usepackage{amsfonts}
\usepackage{graphicx,algorithm,algorithmic,amsmath}
\usepackage{multirow}
\usepackage{array}
\newcommand{\stackcell}[2][c]{%
  \begin{tabular}{@{}#1@{}}
  #2
  \end{tabular}%
}
\usepackage[noadjust]{cite}
\usepackage{subcaption}
\usepackage{color}
\usepackage{flushend}

\title{Second Order Cone Programming for Sensor Node Localization in Mixed LOS/NLOS Conditions}
\author{Sudhir~Kumar,~\IEEEmembership{Student Member,~IEEE,}
       Rishabh~Dixit
        and~Rajesh~M.~Hegde,~\IEEEmembership{Member,~IEEE}
\thanks{The authors are with the Department
of EE, Indian Institute of Technology, Kanpur,
India, e-mail: (sudhirkr@iitk.ac.in, rishabd@iitk.ac.in and rhegde@iitk.ac.in).}}
\begin{document}

\maketitle

\begin{abstract}
In this paper, a novel method for sensor node localization under mixed line-of-sight/non-line-of-sight (LOS/NLOS) conditions based on second order cone programming (SOCP) is presented. 
SOCP methods have, hitherto, not been utilized in the node localization under mixed LOS/NLOS conditions. 
Unlike semidefinite programming (SDP) formulation, SOCP is computationally efficient for resource constrained ad-hoc sensor network. 
The proposed method can work seamlessly in mixed LOS/NLOS conditions.
The robustness of the method is due to the fair utilization of all measurements obtained under LOS and NLOS conditions. 
The computational complexity of this method is quadratic in the number of nearest neighbours of the unknown node.
Extensive simulations and real field deployments are used to evaluate the performance of the proposed method. The experimental results of the proposed method is reasonably better when compared to similar methods in literature. 
\end{abstract}


\section{Introduction}
Localization is a crucial aspect in vehicular sensor networks, where link instability among sensor nodes in NLOS conditions poses a significant challenge. 
The problem of localization using optimization techniques is generally non-linear and non-convex in nature. 
Hence, a closed form solution to the localization problem is not always achievable. 
The problem can be relaxed using semidefinite programming (SDP) or second order cone programming (SOCP) into a convex formulation. 
The sensor node localization problem using SDP and SOCP has been extensively dealt with in  \cite{6847233,chen2012non,lui2009semi,biswas2006semidefinite,tseng2007second} and   \cite{6684554,6847233,srirangarajan2008distributed} respectively. A relation between SOCP and SDP formulation is analysed in \cite{6684554}. It is shown here that SDP provides a tighter relaxation for the sensor node localization problem when compared to SOCP. In the presence of noise, both SOCP and SDP relaxation provide suboptimal solutions \cite{6684554}. There is a trade-off between SDP and SOCP in terms of computational complexity and localization error \cite{tseng2007second}. Notably, the efficacy of solving the localization problem using SOCP relaxation lies in least computational complexity than the counterpart SDP \cite{6684554}. The localization problem using SOCP can be solved in polynomial time. Hence, for tiny sensor nodes with resource constraints, it is imperative to choose a computationally power efficient method. 

A classical SOCP formulation for distributed sensor node localization is proposed in \cite{srirangarajan2008distributed,tseng2007second}. The extension of distributed SOCP (D-SOCP) method  \cite{srirangarajan2008distributed} to account for NLOS conditions increases the localization inaccuracy of sensor node. SOCP method for sensor network localization with anchor uncertainty is suggested in \cite{6684554}. However, though this method considers anchor uncertainty, it fails to capture NLOS conditions. This often occur in practice. A non-cooperative and cooperative localization based on SOCP and SDP relaxation without any explicit assumption of NLOS is discussed in \cite{6847233}. SDP approach for sensor network localization with noisy distance measurements is presented in  \cite{biswas2006semidefinite}. However, \cite{biswas2006semidefinite} considers only measurement noise in distance measurement.  
SDP algorithm for sensor node localization \cite{lui2009semi} is limited to uncertainties in anchor positions and propagation speed. This method does not consider the case of NLOS positive bias. The problem of localization under NLOS condition is addressed in \cite{chen2012non}. This method is based on SDP relaxation, which is computationally expensive compared to SOCP. 

In this work, we propose a SOCP based relaxation method for sensor node localization in Mixed LOS/NLOS conditions (MLN-SOCP). Non-linear least square \cite{lui2009semi} and approximate maximum likelihood \cite{chen2012non} objective functions are exploited in the context of SDP formulation. Subsequently, a NLOS localization problem is formulated in terms of SOCP \cite{6847233, 6684554, srirangarajan2008distributed} for lower computational complexity. 
The proposed method is shown to work in polynomial time. 
Other major advantages of proposed method are its ability to work in mixed LOS/NLOS conditions \cite{yu2009statistical,trees2001detection,kay1998fundamentals}, and robustness to noise under random node deployment scenario.  
The proposed method provides fairness to all distance measurements obtained from both LOS and NLOS. In a large sensor network, there is a scarcity of anchors and obtaining the measurements may become expensive, if NLOS measurements are discarded. 

The notations used in this paper are as follows. The bold faced letter (upper or lower case) represents the matrix. $(.)^T$ denotes the transpose of a matrix. The Euclidean distance between two one-dimensional vectors is represented by $l_2$ norm, $||.||$. The cardinality of a set is denoted by $|.|_c$. The radio communication range of a node is $R$ in $N_d$ $\times$ $N_d$ network. 
The rest of the paper is organized as follows. Section II describes the SOCP formulation for sensor node localization problem. Performance evaluation is demonstrated in Section III, and a brief conclusion is presented in Section IV. Solution to the localization problem in SEDUMI form is listed in the Appendix. 

\section{SOCP Method for Node Localization in Mixed LOS/NLOS Conditions}
In this Section, problem formulation for the sensor node localization is described first. A SOCP relaxation is described next. Subsequently, computational complexity and convergence analysis of the proposed method are presented.   

\subsection{Problem Formulation}
The sensor network is modelled as undirected topology, $\mathcal{G}$ = $(\mathbf{\Xi}, \mathcal{{E}}, \mathbf{\varpi})$. The set of nodes including anchors is represented by $\mathbf{\Xi}$ = $\{1,2, \hdots, m_u, \hdots, p_a\}$. The first $m_u$ nodes are unknown sensor nodes, $\mathbf{\Xi}_u$, while next $p_{a}-m_u$ are anchors or known sensor nodes, $\mathbf{\Xi}_a$. The coordinates of the unknown node and anchor are denoted by $\mathbf{x}_r$ $\in$ $\mathbf{R}^d$ ($d$ = 2 or 3) and  $\mathbf{a}_t$ $\in$ $\mathbf{R}^d$, $\forall$ $r,t$ respectively. The set of edges, $\mathcal{{E}}$ = $\mathcal{{E}}_l$ $\cup$ $\mathcal{E}_n$ among nodes is assigned a weight, $\varpi_{r,t}$ $\in$ $\varpi$ called link quality . 
$\mathcal{E}_l$ and $\mathcal{E}_n$ represent the set of LOS and NLOS edges respectively. 
The channel is assumed to be invariant in both the forward and reverse direction i.e. $\varpi_{r,t}$ = $\varpi_{t,r}$ .

The line-of-sight (LOS) distance between receiving node and emitting node (anchor or known node) is given by 
\begin{equation}
\label{losrange}
d_{r,t}^{l} = ||\mathbf{x}_r - \mathbf{a}_t|| + w_{r,t}, (r,t) \in \mathcal{E}_l,  r<t,
\end{equation}
where $w_{r,t}$ $\sim$ $\mathcal{N}(0,\sigma_{r,t}^{2})$ is the measurement error, which follows a zero-mean Gaussian distribution with variance $\sigma_{r,t}^{2}$. 

The non-line-of-sight (NLOS) distance between receiving node and anchor is corrupted by a large positive error, $o_{r,t}$ due to obstruction. 
\begin{equation}
\label{nlosrange}
\delta_{r,t}^{n} = ||\mathbf{x}_r - \mathbf{a}_t|| + o_{r,t} + w_{r,t},  (r,t) \in \mathcal{E}_n,  r<t,
\end{equation}
where $o_{r,t}$ follows a exponential distribution with mean $\mu_{r,t}$ and variance $\mu_{r,t}^2$ \cite{1595467, yu2008improved,  chen2012non}. A similar model in the context of semidefinite programming is described in \cite{chen2012non}.  The NLOS distance is simplified by subtracting the mean, assuming known NLOS state \cite{chen2012non}.  Consider $\gamma_{r,t}^2$ = $\mu_{r,t}^2$ + $\sigma_{r,t}^{2}$,
\begin{equation}
\label{nlosrange1}
d_{r,t}^{n} = \delta_{r,t}^{n} - \mu_{r,t} = ||\mathbf{x}_r - \mathbf{a}_t|| + w'_{r,t},  (r,t) \in \mathcal{E}_n,  r<t,
\end{equation}
where $w'_{r,t}$ is approximately modelled as Gaussian random variable, $\mathcal{N}(0, \gamma_{r,t}^2)$ \cite{chen2012non}. For large error, this approximation may suffer from the long tail problem. However, it is valid for the most of the scenarios as described in \cite{chen2012non}. 
The joint distribution of LOS and NLOS measurement distances is given by 
\begin{equation}
\begin{aligned}
\label{jointpdf}
\mathscr{P}(d_{r,t}^{l}, d_{r,t}^{n}|\mathbf{\Xi}_{u}) = \prod_{(r,t) \in \mathcal{E}; r<t} [g_{r,t}\mathcal{P}_{w}\{(d_{r,t}^{l} - ||\mathbf{x}_r - \mathbf{a}_t||)\\|\text{LOS}\}  + (1 - g_{r,t})\mathcal{P}_{w'}\{(d_{r,t}^{n} - ||\mathbf{x}_r - \mathbf{a}_t||)|\text{NLOS}\}],
\end{aligned}
\end{equation}
where $\mathcal{P}_{w}$ and $\mathcal{P}_{w'}$ are the probability density function of the error $w_{r,t}$ and $w_{r,t}'$. The probability of LOS link is denoted by $g_{r,t}$. 
Following \cite{chen2012non}, the objective function using the approximate maximum likelihood method can be recast as 
\begin{equation}
\label{objaml}
\begin{aligned}
 \hspace{-10mm}\underset{\mathbf{\Xi}_u}{\text{minimize}}
 \sum\limits_{(r,t) \in \mathcal{E};r<t} \bigg[\frac{1}{\sigma_{r,t}^{2}}g_{r,t}\left|||\mathbf{x}_r - \mathbf{a}_t|| - d_{r,t}^{l}\right|^2 \\  + \frac{1}{\gamma_{r,t}^{2}}(1- g_{r,t})\left|||\mathbf{x}_r - \mathbf{a}_t|| - d_{r,t}^{n}\right|^2 \bigg]
\end{aligned}
\end{equation}
The variance of LOS link is $\sigma_{r,t}^{2}$ = $\eta_l^2(d_{r,t}^{l})^2$, while variance of NLOS link is $\gamma_{r,t}^{2}$ = $(\eta_l^2 + \eta_n^2)(d_{r,t}^{n})^2$. Standard deviation per unit length of noise and NLOS error are  represented by $\eta_l$ and $\eta_n$ respectively. 
In order to obtain the location of unknown sensor nodes, Equation \ref{objaml} needs to be minimized over the entire network. 
This optimization problem is non-linear in nature and it is difficult to find a closed loop solution. Finding the global optima is also not easy, because the problem is non-convex in nature. 
\subsection{Solution using SOCP Relaxation} 
To solve this optimization problem, second order cone programming (SOCP) relaxation is used, which is computationally efficient for sensor node localization.  
Defining auxiliary variables $\{q_{r,t}\}$ and $\{z_{r,t}\}$,
the problem in Equation \ref{objaml} can be reformulated using relaxed conic constraints as 
\begin{equation}
\begin{aligned}
& \underset{\mathbf{\Xi}_u, \{q_{r,t}\}, \{z_{r,t}\}}{\text{minimize}}
& & \sum\limits_{(r,t)\in \mathcal{E};r<t} (q_{r,t}^2 + z_{r,t}^2) \\
& \text{subject to} 
& & \sqrt{\frac{g_{r,t}}{\sigma_{r,t}^{2}}}\left|||\mathbf{x}_r - \mathbf{a}_t|| - d_{r,t}^{l}\right| \le q_{r,t}, \; (r,t) \in \mathcal{E}_l \\
&&& \sqrt{\frac{1-g_{r,t}}{\gamma_{r,t}^{2}}}\left|||\mathbf{x}_r - \mathbf{a}_t|| - d_{r,t}^{n}\right| \le z_{r,t}, \; (r,t) \in \mathcal{E}_n
\end{aligned}
\end{equation}
Introducing an epigraph variable $V$ \cite{boyd2009convex, 6684554} and auxiliary variables $\mathbf{U}$ and $\{y_{r,t}\}$, the equivalent convex epigraph problem is written  as
\begin{equation}
\label{final_equation}
\begin{aligned}
& \hspace{-12mm} \underset{\mathbf{\Xi}_u, \{q_{r,t}\}, \{z_{r,t}\}, \{y_{r,t}\}, V}{\text{minimize}}  V \\
 \text{subject to} & \\
& ||\mathbf{U}|| \le V\\
& \sqrt{\frac{g_{r,t}}{\sigma_{r,t}^{2}}}\left|y_{r,t} - d_{r,t}^{l}\right| \le q_{r,t}, \; (r,t) \in \mathcal{E}_l \\
& \sqrt{\frac{1-g_{r,t}}{\gamma_{r,t}^{2}}}\left|y_{r,t} - d_{r,t}^{n}\right| \le z_{r,t}, \; (r,t) \in \mathcal{E}_n \\
& ||\mathbf{x}_r - \mathbf{a}_t|| \le y_{r,t}
\end{aligned}
\end{equation}
where $\mathbf{U}$ = $[q_{r,t} \hspace{2mm} z_{r,t}]$. Fourth constraint is induced from the  relaxation of equality constraint into an inequality one. 
Localization of the sensor node is now reduced to solving Equation \ref{final_equation}, which represents a SOCP relaxation of original non-convex optimization problem. The solution of Equation \ref{final_equation} is obtained by numerical optimization technique like SEDUMI solver \cite{sturm1999using}. The solution\footnotemark                                     is enumerated in Appendix.
\footnotetext[1]{Solution to the problem in SEDUMI form is listed in the Appendix}
\begin{figure}[!ht] 
  \begin{subfigure}[b]{0.5\linewidth}
    \centering
    \includegraphics[height=40mm,width=48mm]{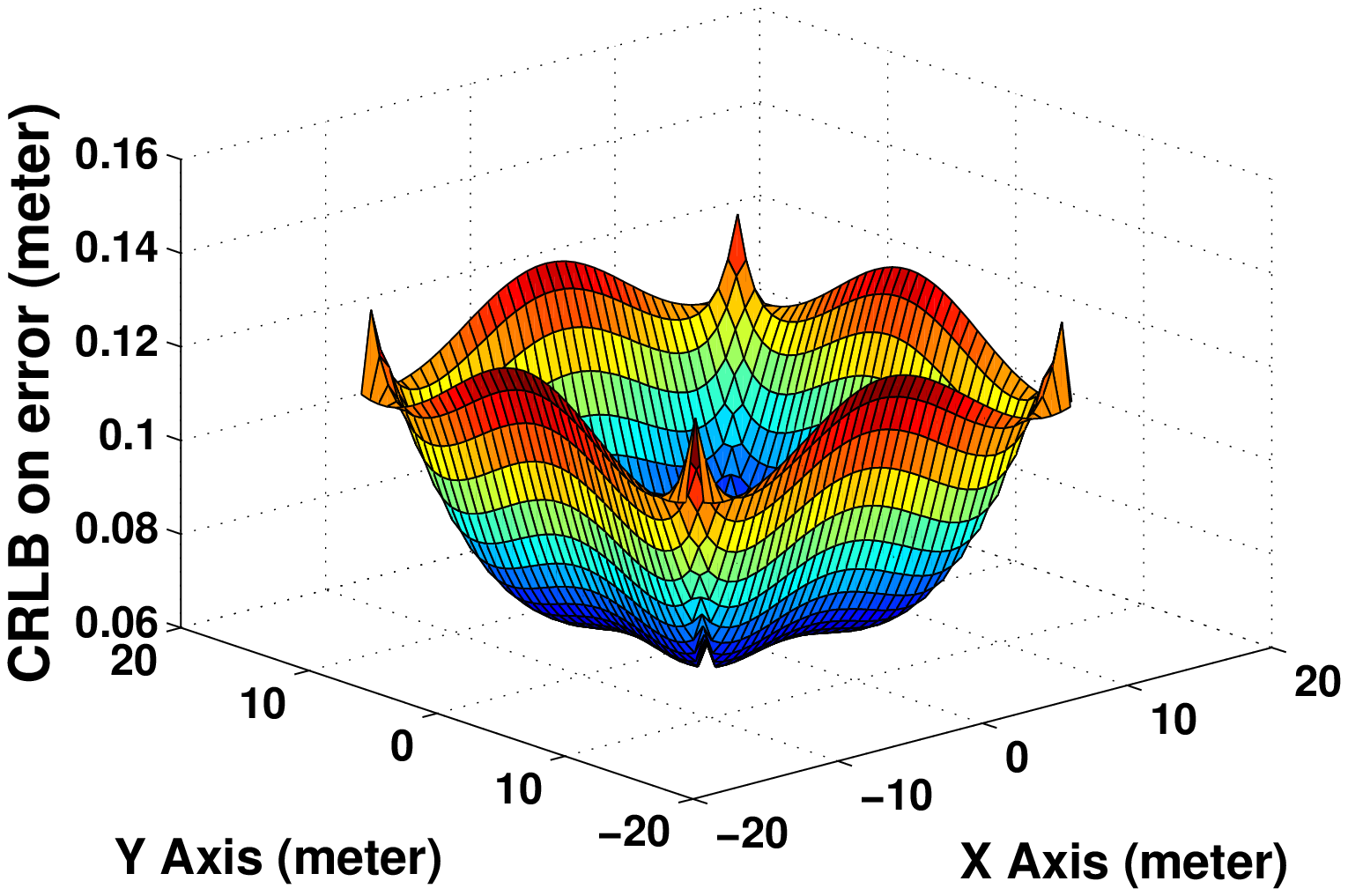} 
  \end{subfigure}
  \begin{subfigure}[b]{0.5\linewidth}
    \centering
    \includegraphics[height=40mm,width=48mm]{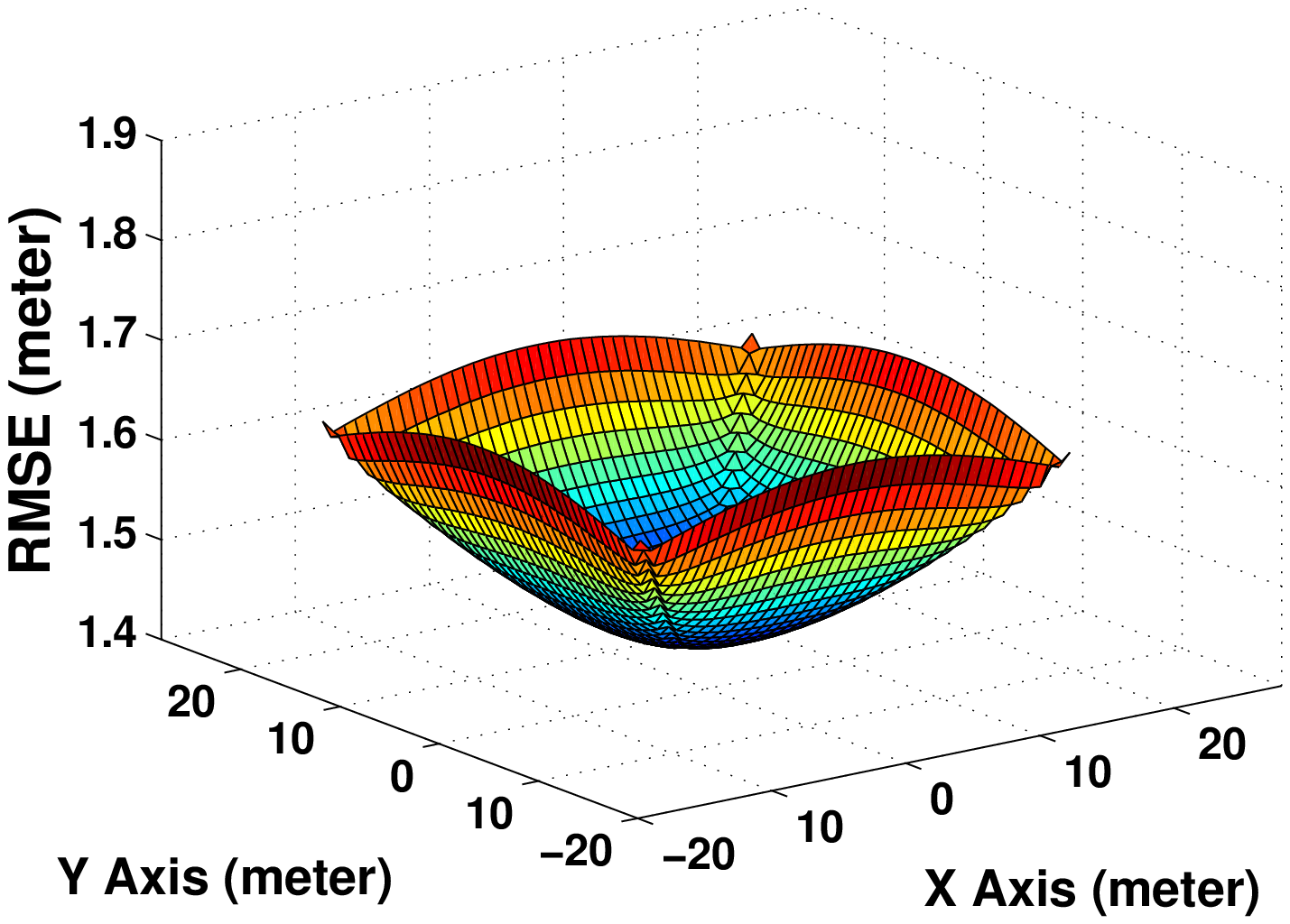} 
  \end{subfigure} 
  \caption{Figure illustrating the (a) Cram{\'e}r-Rao lower bound and (b) Root mean squared error plot for the proposed method in a 40 m $\times$ 40 m network; for $p$ = 0.0025, $g_{r,t}$ = 0.7, $|\mathbf{\Xi}|_c$ = 1600, $R$ = $N_d$, $\eta_l$ = 0.1, $\eta_n$ = 0.06.}
  \label{crlb} 
\end{figure}

\subsection{Computational Complexity Analysis}
Let $i^{th}$ unknown sensor node can be localized using $p_i$ number of neighbouring anchors within its vicinity, which are contained in set $P_i$. The computational complexity \cite{polik2010interior, 6847233} of the proposed method is given by 
\begin{equation}
\label{computational_complexity}
\mathcal{O}\bigg(M(e^3 + \sum_{j=1}^{N} k_j^2 + e^2\sum_{j=1}^{N} k_j)\bigg)
\end{equation}
where $e$ and $M$ denote the total number of equality constraints and iteration complexity respectively. $k_j$ and $N$ represent the dimension of $j^{th}$ second order cone and total number of second order cone constraints respectively.  The total number of conic constraints is $3p_{i}+1$, $\forall$ $t$ as can be noted from Equation \ref{final_equation}. The number of equality constraint for the SOCP formulation in Equation \ref{final_equation} is zero. The dimension of first to fourth set of constraints are $(2p_i+1)$, $2$, $2$ and $3$ respectively. Therefore, asymptotic computational complexity for the proposed MLN-SOCP method is reduced to $\mathcal{O}(M|P_i|_c^2)$, which is quadratic in cardinality of $P_i$. Computational complexity can be reduced by constraining $|P_i|_c$ to localize an unknown sensor node. 

%

\section{Performance Evaluation}
In this Section, experimental conditions for real field deployment is presented first. Subsequently, experimental results for sensor node localization in mixed LOS/NLOS are discussed.
\subsection{Experimental Conditions}
An experimental ad-hoc sensor network is deployed using both National Instrument (NI) WSN node and Crossbow motes to validate the effectiveness of the proposed algorithm. 
For a larger dimensions of the network, NI WSN - 3202, 3212 and gateway NI 9792 are used. 
For a smaller network dimensions, Crossbow MTS310 sensor board, MIB520 USB mote interface board and XM2110 IRIS board are used. 
The nodes communicate among themselves through IEEE 802.15.4 protocol.
Sensor nodes are randomly deployed inside the network assuming a uniform distribution. 

\subsection{Experimental Results}
In this Section, Cram{\'e}r-Rao lower bound analysis is illustrated first. Subsequently, localization error analysis followed by probabilistic error analysis of location estimation are presented.
 
\subsubsection{Cram{\'e}r-Rao Lower Bound Analysis}
In order to assess minimum variance of the localization error, Cram{\'e}r-Rao lower bound \cite{patwari2003relative, 6831118} analysis plot is illustrated in Figure \ref{crlb}. For sake of illustration, anchors are placed at the boundary of the network, while nodes are deployed inside the convex hull of networks. The minimum CRLB on localization error is found to be $0.073$, which is at the center of the surface.
The corresponding root-mean-square error (RMSE) plot using estimated distance from the proposed algorithm is shown in Figure \ref{crlb}(b). The minimum value attained for the proposed method, MLN-SOCP is $1.4719$, while for D-SOCP method \cite{srirangarajan2008distributed} is $3.2682$. 

\subsubsection{Localization Error Analysis}
In this Section, variation of localization error with various network parameters namely network dimension, $N_d$, radio communication range, $R$, fraction of anchors, $p$ and probability of LOS links, $g_{r,t}$ are illustrated. Localization error is defined as the Euclidean distance between actual and estimated node location. 

\textit{(a) Comparison of Localization Error with Varying Network Dimension and Radio Communication Range:} 
If the anchors are linearly separable over all nodes in the networks, localization error  \cite{nguyen2005kernel} is given by 

\begin{equation}
\label{errorBound}
\mathbb{E}||\mathbf{x}_r - \mathbf{\hat{x}}_r|| = \mathcal{O}(N_d^{\frac{1}{3}}R^{\frac{2}{3}}{|\mathbf{\Xi}|_a}^{\frac{-1}{6}})
\end{equation}

where network is considered to be of size $N_d$ $\times$ $N_d$. The communication range of the node  is $R$, while ${|\mathbf{\Xi}|_a}$ represents the cardinality of set of anchors. $\mathbb{E}$ represents the  expectation operator. 
As the network dimension increases, localization error increases as shown in Table \ref{table1}. The proposed method, MLN-SOCP performs significantly better than the D-SOCP for all values of radio communication range, $R$. The increase in the localization error with network dimension can be at-most linear \cite{nguyen2005kernel}. 
\begin{table}[h]
\centering
\caption{ Comparison of localization error (m) for various $N_d$, $R$, $\eta_l$ and $\eta_n$; for $p$ = 0.3, $g_{r,t}$ = 0.7, $|\mathbf{\Xi}|_c$ = 100.}
\scalebox{0.85}{
\begin{tabular}{|c|c|c|c|c|c|}
\hline            
   \textbf{Methods}  & 
\bfseries\stackcell{Radio} & \multicolumn{2}{|c|}{${\boldsymbol{\eta_l}}$ = \textbf{0.10}, $\boldsymbol{\eta_n}$ = \textbf{0.06}} & \multicolumn{2}{|c|}{\bfseries \stackcell{$\mathbf{N_d}$ = $\mathbf{40m}$}} \\  \cline{3-3} \cline{4-4} \cline{5-5} \cline{6-6} 
& \bfseries{Range} & $\mathbf{N_d}$ = $\mathbf{40m}$ & $\mathbf{N_d}$ = $\mathbf{80m}$ & \stackcell{$\boldsymbol{\eta_l}$ = \textbf{0.2}\\$\boldsymbol{\eta_n}$ = \textbf{0.15}} & \stackcell{$\boldsymbol{\eta_l}$ = \textbf{0.3}\\ $\boldsymbol{\eta_n}$ = \textbf{0.25}}\\
\hline \hline
  \bfseries {D-SOCP} & $\sqrt{2}N_d$ &
4.30 &	10.05	&  6.99 & 8.40
 \\  \cline{2-2} \cline{3-3}
\cline{4-4} \cline{5-5} \cline{6-6}
 & $N_d$ & 3.99 &	8.34	& 6.97 &  7.90 
 \\ \cline{1-1} \cline{2-2} \cline{3-3} \cline{4-4} \cline{5-5} \cline{6-6}
\textbf{MLN-SOCP} 
 & $\sqrt{2}N_d$ & 3.84 &	8.24 &	4.40 & 6.52 
 \\  \cline{2-2} \cline{3-3} \cline{4-4} \cline{5-5} \cline{6-6}

 & $N_d$& 2.83 &	4.27 &	4.39 & 6.33

 \\ \cline{1-1}
\cline{2-2} \cline{3-3} \cline{4-4} \cline{5-5} \cline{6-6}
\hline 
\end{tabular}}
\label{table1}
\end{table}
However, if the ratio of network dimension to the number of anchors is kept constant, then it is likely that localization error is unaffected. On the other hand, if the number of anchors are held constant, there is an increase of localization error. This is due to the fact that the node may not able to hear sufficient number of anchors for the localization. Increase in  radio communication range enables a node to hear many anchors. But, incorporating many anchors increases the possibility of inclusion of erroneous distances between anchor and node. This results in an  increase of the localization error at the rate of $\mathcal{O}(R^{\frac{2}{3}})$  \cite{nguyen2005kernel}. There is a marginal increase in localization error with standard deviation of noise and NLOS. 

\textit{(b) Comparison of Localization Error with Varying LOS Probability and Node Density:} As the node density increases, the localization error reduces as expected in Table \ref{table_nodedensity}. However, localization error decreases till a certain $\mathbf{|\Xi}|_c$. Increasing the node density implies an increase in number of anchors by a proportion '$p$'. Inclusion of more anchors leads to the increase of erroneous distance between unknown node and anchor. This results in reduced localization accuracy after, $|\mathbf{\Xi}|_c$ = 200. There is a marginal decrease in localization error on increasing the LOS probability links, $g_{r,t}$ as expected. 
 
\begin{table}[h]
\centering
\caption{ Comparison of localization error (m) for varying $\mathbf{|\Xi}|_c$ and $g_{r,t}$; for $p$ = 0.3, $R$ = $N_d$ = $40m$, $\eta_l$ = 0.1, $\eta_n$ = 0.06.}
\scalebox{0.9}{
\begin{tabular}{|c|c|c|c|c|c|c|c|}
\hline            
\multicolumn{2}{|c|}{} & \multicolumn{6}{c|}{\bfseries{$\mathbf{|\Xi}|_c$}} \\
\hline
   \textbf{Methods}  & $\mathbf{g_{r,t}}$ & 
\textbf{50} & \textbf{100} & \textbf{150} & \textbf{200} & \textbf{250} & \textbf{300} \\
\hline \hline
   & 0.95 &  4.05	& 3.70 &	3.66	& 3.54 &	3.91	& 4.01

 \\  \cline{2-2} \cline{3-3}
\cline{4-4} \cline{5-5} \cline{6-6} \cline{7-7} \cline{8-8}
   & 0.7 & 4.36	& 3.99 & 3.78 &	3.65 &	4.09 &	4.49

 \\  \cline{2-2} \cline{3-3}
\cline{4-4} \cline{5-5} \cline{6-6} \cline{7-7} \cline{8-8}
 \bfseries {D-SOCP}  & 0.4 & 5.68	& 5.52 &	5.16 &	5.11 &	5.27 & 	5.73

 \\  \cline{2-2}  \cline{3-3} \cline{4-4} \cline{5-5}
\cline{6-6} \cline{7-7} \cline{8-8}
& 0.1 & 5.79 &	5.56	& 5.45 &	5.25 &	5.31 &	5.88

  \\  \cline{1-1} \cline{2-2}  \cline{3-3} \cline{4-4} \cline{5-5}
\cline{6-6} \cline{7-7} \cline{8-8}
\cline{6-6} \cline{7-7} \cline{8-8}
 & 0.95 &  2.64	& 2.55 &	2.48 &	2.39	& 2.44 &	2.53

 \\  \cline{2-2} \cline{3-3}
\cline{4-4} \cline{5-5} \cline{6-6} \cline{7-7} \cline{8-8}
   & 0.7 & 2.85	& 2.83 &	2.54 &	2.50 &	2.74 &	2.77

 \\  \cline{2-2} \cline{3-3}
\cline{4-4} \cline{5-5} \cline{6-6} \cline{7-7} \cline{8-8}
 \bfseries {MLN-SOCP}  & 0.4 & 2.87	& 2.85 & 2.60 &	2.59 &	2.81 & 	2.82

  \\  \cline{2-2}  \cline{3-3} \cline{4-4} \cline{5-5}
\cline{6-6} \cline{7-7} \cline{8-8}
& 0.1 & 2.91	& 2.90 &	2.69 &	2.68 &	2.84 &	2.86

  \\  \cline{2-2}  \cline{3-3} \cline{4-4} \cline{5-5}
\cline{6-6} \cline{7-7} \cline{8-8}
\hline 
\end{tabular}}
\label{table_nodedensity}
\end{table}

\subsubsection{Probabilistic Error Analysis of Location Estimation}
\begin{figure}[h]
\begin{center}
 \includegraphics[height=50mm,width=80mm]{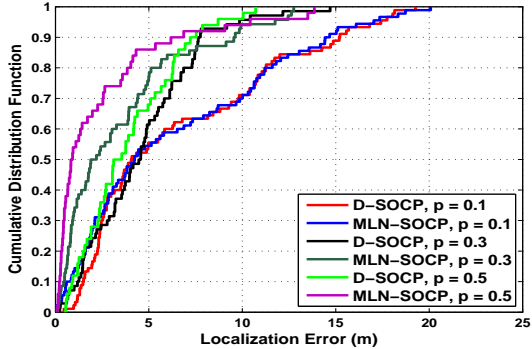}
 \caption{Figure illustrating the CDF plot for various $p$; for $g_{r,t}$ = 0.7, $|\mathbf{\Xi}|_c$ = 100, $R$ = $N_d$ = $40m$, $\eta_l$ = 0.1, $\eta_n$ = 0.06.}
 \label{cdf}
\end{center}
 \end{figure}

The cumulative distribution function (CDF) represents the statistical distribution of localization error.  Figure \ref{cdf} shows the empirical CDF computed for various values of $p$. Since the proposed MLN-SOCP method uses the maximum likelihood model, the MLN-SOCP performs significantly better than the D-SOCP method for higher values of probability of anchors. This aspect can be verified from Figure \ref{cdf}, since the CDF curve rapidly attains a probability of unity. However, for lower values of $p$ = $0.1$, MLN-SOCP method performs reasonably better than D-SOCP.    
\section{Conclusion}
In this paper, second order cone programming for sensor node localization in mixed LOS/NLOS is  proposed. 
The novelty of the proposed method lies in the development of NLOS model using SOCP relaxation. 
The experimental results on sensor node localization in mixed LOS/NLOS illustrate that the node can be localized with a high degree of reliability even for large number of NLOS links using the  maximum likelihood model. 
The results also indicate that the proposed method can be effectively used for node tracking in vehicular ad-hoc networks.  
New methods that can utilize SOCP to detect and localize malicious anchors are currently being explored.  
\section{Appendix}
Let $r^{th}$ unknown sensor node can be localized using $p_i$ number of neighbouring anchors within its vicinity. 
\begin{equation}
\label{ith_sensor_neighbour}
\mathbf{\Xi}_r = \begin{bmatrix} x_r & y_r & x_1 & x_2 & \hdots & x_{p_{i}} & y_{p_{i}}\end{bmatrix}^T_{(2p_{i}+2) \times 1}
\end{equation}
The objective function of Equation \ref{final_equation} can be written in SEDUMI form as 
\begin{equation}
\label{xbar}
\mathbf{\tilde{x}} = \begin{bmatrix} \mathbf{\Xi}_r^T & \mathbf{q}_{r,t} & \mathbf{z}_{r,t} & \mathbf{y}_{r,t} & V & \end{bmatrix}^T
\end{equation}
where
\begin{equation}
\begin{aligned}
\label{qrt_zrt_yrt}
& \mathbf{q}_{r,t} = \begin{bmatrix} q_{r,1} & q_{r,2} & \hdots & q_{r,p_{i}}  \end{bmatrix}_{1 \times p_{i}} \\
& \mathbf{z}_{r,t} = \begin{bmatrix} z_{r,1} & z_{r,2} & \hdots & z_{r,p_{i}}  \end{bmatrix}_{1 \times p_{i}} \\
& \mathbf{y}_{r,t} = \begin{bmatrix} y_{r,1} & y_{r,2} & \hdots & y_{r,p_{i}}  \end{bmatrix}_{1 \times p_{i}}
\end{aligned}
\end{equation}
The objective function in the SEDUMI is of the form 
\begin{equation}
\label{BTx}
\mathbf{B}^T\mathbf{\tilde{x}} = V, \text{where } \mathbf{Bt} = -\mathbf{B} = - \begin{bmatrix} \mathbf{0}_{(1,5p_{i}+2)} & 1   \end{bmatrix}^T
\end{equation}
The total number of constraints are $3p_{i}+1$. The dimension of first to fourth set of constraints are $2p_i+1$, $2$, $2$ and $3$ respectively.
Comparing with standard form of conic constraint,
\begin{equation}
\label{std_conic}
||\mathbf{A}^T\mathbf{\tilde{x}} + \mathbf{c}|| \le \mathbf{b}^T\mathbf{\tilde{x}} + \mathbf{d}
\end{equation}
For first constraint in Equation \ref{final_equation}, 
\begin{equation}
\label{constraint1}
\mathbf{A}_1^T\mathbf{\tilde{x}} = \mathbf{U}, \mathbf{c}_1 =0, \mathbf{b}_1^T\mathbf{\tilde{x}} = V \hspace{1mm}\text{and}\hspace{1mm} \mathbf{d}_1 =0
\end{equation}
Therefore,
\begin{equation}
\label{Ai}
\mathbf{A}_1 = \begin{bmatrix}  \mathbf{E}_{1} & \mathbf{E}_{2} & \hdots & \mathbf{E}_{k} & \dots & \mathbf{E}_{p_i} \end{bmatrix}
\end{equation}
where 
\begin{equation}
\label{Ai}
\mathbf{E}_k = \begin{bmatrix}  \mathbf{E}_{k,1} & \mathbf{E}_{k,2}  \end{bmatrix}
\end{equation}
\begin{equation}
\begin{aligned}
\label{stackofA}
&\mathbf{E}_{k,1} = \begin{bmatrix} \mathbf{0}_{(1,2p_{i}+2)} & \mathbf{1}_{k,t} &  \mathbf{0}_{(1,2p_{i}+1)} \end{bmatrix}^T \\ 
&\mathbf{E}_{k,2} = \begin{bmatrix} \mathbf{0}_{(1,3p_{i}+2)} & \mathbf{1}_{k,t} &  \mathbf{0}_{(1,p_{i}+1)} \end{bmatrix}^T \\
&\mathbf{b}_{1} = \begin{bmatrix} \mathbf{0}_{(1,5p_{i}+2)} & 1  \end{bmatrix}^T 
\end{aligned}
\end{equation}
with $\mathbf{1}_{k,t}$ is row vector of length $p_i$ with a 1 corresponding to the index $k$ and 0's elsewhere. 
Similarly, for second set of constraints
\begin{equation}
\begin{aligned}
\label{stackofsecondconstraint}
&\mathbf{A}_{2,k} = \begin{bmatrix} \mathbf{0}_{(1,4p_{i}+2)} & \mathbf{1}_{k,t} &  0 \end{bmatrix}^T \\ 
&\mathbf{b}_{2,k} = \frac{\eta_l}{\sqrt{g_{k,t}}}d_{k,t}^{l}\begin{bmatrix} \mathbf{0}_{(1,2p_{i}+2)} & \mathbf{1}_{k,t} &  \mathbf{0}_{(1,2p_{i}+1)} \end{bmatrix}^T \\
&\mathbf{c}_{2,k} = \begin{bmatrix} 0 & -d_{k,t}^{l}  \end{bmatrix}, 
\mathbf{c}_{2} = \begin{bmatrix} \mathbf{c}_{2,1} & \hdots & \mathbf{c}_{2,p_i}  \end{bmatrix} \\
& \mathbf{A}_2 = \begin{bmatrix} \begin{bmatrix} \mathbf{b}_{2,1} & \mathbf{A}_{2,1} \end{bmatrix} & 
\hdots &
\begin{bmatrix} \mathbf{b}_{2,p_i} & \mathbf{A}_{2,p_i} \end{bmatrix} \end{bmatrix} 
\end{aligned}
\end{equation}
On similar line, matrices for third set of constraints can be written. 
For fourth set of constraints, 
\begin{equation}
\begin{aligned}
\label{stackofsecondconstraint}
&\mathbf{A}_{4,k} = \begin{bmatrix} 1  &  0 & \mathbf{0}_{(1,5p_{i}+1)} \\ 0  &  1 & \mathbf{0}_{(1,5p_{i}+1)} \end{bmatrix}^T \\ 
& \mathbf{b}_{4,k} = \begin{bmatrix} \mathbf{0}_{(4p_{i}+2, 1)} & \mathbf{1}_{k,t}^T & 0 \end{bmatrix}^T, 
\mathbf{c}_{4,k} = \begin{bmatrix} 0 & -\mathbf{a}_t^T \end{bmatrix}
\end{aligned}
\end{equation}
Similarly, $\mathbf{A}_4$ and $\mathbf{c}_4$ can be written like  $\mathbf{A}_2$ and $\mathbf{c}_2$.
\begin{equation}
\label{finalA}
\mathbf{A} = -\begin{bmatrix} \begin{bmatrix} \mathbf{b}_1 & \mathbf{A}_1 \end{bmatrix}  &  
\mathbf{A}_2
&\hdots  
&  \mathbf{A}_4  \end{bmatrix}
\end{equation}
\begin{equation}
\label{finalC}
\mathbf{C} = \begin{bmatrix} \mathbf{0}_{1,p_{i}} & \mathbf{c}_2 & \mathbf{c}_3  & \mathbf{c_4} \end{bmatrix}^T
\end{equation}
Cartesian product of all cones which is the dimension of constraints 
\begin{equation}
\label{finalC}
\mathbf{Q} = \begin{bmatrix} 2p_{i}+1 & 2.\mathbf{1}_{(1,2p_i)} & 3.\mathbf{1}_{(1,p_i)}   \end{bmatrix}
\end{equation}
Hence, Equation \ref{final_equation} can be written in terms of  $\mathbf{A}$, $\mathbf{Bt}$ , $\mathbf{C}$ and $\mathbf{Q}$. 
\newpage
\bibliographystyle{IEEEtran}
\bibliography{References}
\end{document}